\documentclass[12pt,nohyper,notoc]{JHEP}

\usepackage{amsmath,amssymb,cite} 
	
\DeclareSymbolFont{AMSb}{U}{msb}{m}{n}
\DeclareSymbolFontAlphabet{\Bbb}{AMSb}
\def\rsen{\setcounter{equation}{0}}

\newcommand{\startappendix}{
\setcounter{section}{0}
\renewcommand{\thesection}{\Alph{section}}
\renewcommand{\theequation}{\Alph{section}.\arabic{equation}}}

\newcommand{\Appendix}[1]{
\refstepcounter{section}
\begin{flushleft}
{\large\bf Appendix \thesection: #1}
\end{flushleft}}

\newcommand{\Rowspace}{\phantom{$\Big($}}

\newcommand{\aD}{{\dot\alpha}}

\newcommand{\bD}{{\dot\beta}}

\newcommand{\BL}{{\bf L}} 

\newcommand{\A}{{\cal A}}

\def\com{X}
\def\sfc{\hat\Omega}

\def\N{{\cal N}}
\def\det{{\rm det}}

\def\M{{\cal M}}

\def\Mbar{\bar{\cal M}}

\def\Skinst{S^k_{\rm inst}}
\def\dmuphys{d\mu^{k}_{\rm phys}}
\def\Skquad{S^k_{\rm quad}}
\def\sqrtwo{\sqrt{2}\,}

\def\sst{\scriptscriptstyle}

\def\N{{\cal N}}
\def\det{{\rm det}}

\def\SU{\text{SU}}
\def\SO{\text{SO}}
\def\Sp{\text{Sp}}
\def\O{\text{O}}
\def\U{\text{U}}
\def\H{\text{H}}

\title{Instantons in N=4 Sp(N) and SO(N) theories and the AdS/CFT correspondence}

\author{Timothy J.~Hollowood$^{a,c}$, Valentin V.~Khoze$^b$ and
Michael P.~Mattis$^a$\\
$^a$Theoretical Division T-8, Los Alamos National Laboratory,
Los Alamos, NM 87545, USA\\
$^b$Department of Physics, University of Durham,
Durham, DH1 3LE, UK\\
$^c$Department of Physics, University of Wales Swansea,
Swansea, SA2 8PP, UK\\
E-mail: {\tt pyth@skye.lanl.gov}, {\tt valya.khoze@durham.ac.uk}, {\tt
mattis@lanl.gov}} 

\abstract{Following work on theories with $\SU(N)$ gauge
groups, we perform a large-$N$ saddle-point approximation of the measure for ADHM
multi-instantons in ${\cal
N}=4$ supersymmetric gauge theories with symplectic or orthogonal
gauge groups. For $\Sp(N)$ we find that a saddle-point only exists in
the even instanton charge sector. For either $\Sp(N)$ or $\SO(N)$ 
the saddle-point solution parametrizes
$AdS_5\times{\Bbb R}P^5$, the dual supergravity geometry in the 
AdS/CFT correspondence for these theories. The instanton measure
at large-$N$ has the form of the partition function of ten-dimensional
$\N=1$ supersymmetric gauge theory with a unitary gauge group dimensionally
reduced to zero dimensions.}


\preprint{{\tt hep-th/9910118}}

\begin{document}

\section{Introduction}

This paper is concerned with the remarkable series of conjectures
that have been made postulating a
duality in the 't~Hooft coupling of 
conformally invariant gauge theories and type IIB string theory on
a background of the form $AdS_5\times M_5$, where $M_5$ is a certain
five dimensional compact space. In the simplest case---the original
conjecture of Maldacena \cite{MAL}---the gauge theory
has ${\cal N}=4$ supersymmetry with $\SU(N)$ gauge group 
in which case $M_5$ is $S^5$ (see also \cite{GKP,WIT150,AGMOO:rev}). 
When the gauge group is one of the other classical
groups, the five sphere is replaced by the projective space 
$S^5/{\Bbb Z}_2\simeq{\Bbb R}P^5$ \cite{Kakushadze:1998tr,Kakushadze:1998tz}. 
There are many other generalizations; however, in this paper we
concentrate on the $\N=4$ theories with gauge groups $\Sp(N)$ and
$\SO(N)$. In this sense this is a companion to \cite{MO-III} which
considered the $\SU(N)$ theories.

Refs.~\cite{lett,MO-III} uncovered striking relations between the
multi-instantons of $\N=4$ supersymmetric gauge theory and
D-instantons in type IIB supergravity building on the ideas of
\cite{BGKR,Green:rev}. 
The first relation involves the
theory at finite $N$ and the second at large-$N$:

(i) At finite $N$, we showed that the measure for ADHM instantons was
precisely equal to the partition function of D-instantons moving in a
background of $N$ coincident D3-branes in the decoupling limit
$\alpha'\rightarrow0$. This partition function is precisely the dimensional
reduction to zero dimensions of the pure $\N=(1,1)$ supersymmetric $\U(k)$
gauge theory in six dimensions with $N$ additional hypermultiplets (and so the
resulting theory actually only has $\N=(1,0)$ supersymmetry).
What is particularly striking is that the auxiliary scalars $\chi$ introduced 
in \cite{MO-III} to
bi-linearize a certain four-fermion interaction in the instanton
action arises in a very natural way as the scalars corresponding to
the six-dimensional gauge field and describe the freedom for the 
D-instantons to be ejected from the D3-branes.

(ii) At large $N$, the ADHM measure can be
approximated by an expansion around a saddle-point 
and is proportional to
the partition function of the six-dimensional pure $\N=(1,1)$ 
supersymmetric $\U(k)$ gauge theory---with no additional 
matter---dimensionally reduced to zero dimensions.
Alternatively, this can be described as the 
ten dimensional $\N=1$ supersymmetric $\U(k)$ gauge theory dimensionally
reduced to zero dimensions. This is
precisely what one expects for D-instantons in a flat background with
no D3 branes present where the $\U(1)^k\subset\U(k)$ components of the
gauge field are interpreted as
the position of the charge-$k$ D-instanton in ${\Bbb R}^{10}$. 
The only difference with the
large-$N$ measure is that the abelian component of the gauge field is now
interpreted as the position in $AdS_5\times S^5$: the near horizon
geometry of the $N\rightarrow\infty$ D3-branes. 

The large-$N$ approximation of the measure described in (ii) was
performed by a steepest-decent method. This gives a way to probe the
ten-dimensional geometry directly since the solution of the
saddle-point equations could be interpreted as the position of a
point-like object in $AdS_5\times S^5$. From the gauge theory side the
solution represents the simple configuration where all the instantons are at
the same point in ${\Bbb R}^4$ with the same scale size and in
mutually commuting $\SU(2)$ subgroups of the gauge group. The other
five coordinates arise from auxiliary scalar variables that are used
to bi-linearize a certain four-fermion interaction mediated by the
Yukawa couplings in the gauge theory. These additional variables are 
$\SO(6)$, the $R$-symmetry of the $\N=4$ theory, vector-valued and at
the saddle-point they are constrained to lie on $S^5$.  

This qualitative relation between large-$N$ ADHM instantons and
D-instantons in $AdS_5\times S^5$ was made quantitative since
D-instantons contribute terms to the type IIB effective
\cite{BG,BGKR,Green:rev}
action that
imply very particular ADHM instanton contributions to certain
correlation functions. In Ref.~\cite{MO-III} we found precise
agreement between the D-instanton induced effects and ADHM
contributions to the relevant correlation functions.
This paper investigates the extent to which this relation between
D-instantons and ADHM instantons persists when the gauge group of the
theory is either $\SO(N)$ or $\Sp(N)$. Other generalizations involving
$\N<4$ supersymmetric theories either based on orbifolds of the
$\SU(N)$ theory or the finite $\N=2$ $\Sp(N)$ theory have
been considered in Refs.~\cite{orbi,spnt}. We will not consider the
finite-$N$ relation spelled out in (i) above; rather we shall
concentrate on the large-$N$ situation. 
Our results are as follows:

(i) For gauge group $\Sp(N)$ the ADHM construction at charge $k$
involves an auxiliary group $\O(k)$. In this case
the large-$N$ saddle-point equations
only have a solution for even instanton number. The saddle-point
solution in the instanton charge $2k$ sector describes 
the positions of $k$ point-like objects, the D-instantons of the
string theory, in $AdS_5\times{\Bbb R}P^5$. 
The expansion around the general saddle-point solution can be captured
by an expansion around the maximally degenerate solution where all the
D-instantons are at the same point in $AdS_5\times{\Bbb R}P^5$. This
solution is invariant under $\U(k)\subset\O(2k)$. 

(ii)  For gauge group $\SO(N)$ the ADHM construction at charge $k$
involves an auxiliary group $\Sp(k)$. In this case the
large-$N$ saddle-point equations have a solution for all instanton
numbers $k$ which describes the positions of $k$ point-like objects, 
the D-instantons of the
string theory, in $AdS_5\times{\Bbb R}P^5$. In this case the maximally
degenerate saddle-point solution is invariant under $\U(k)\subset\Sp(k)$.

In both cases (i) and (ii) the
large-$N$ ADHM instanton measure has the same form as that of the
instanton charge $k$ sector of the $\N=4$ $\SU(N)$ theory. In other words at
leading order the measure is equal to the partition function of
ten-dimensional $\N=1$ supersymmetric $\U(k)$ dimensionally reduced to
zero dimensions (the unbroken residual
symmetry group in both cases). The abelian components correspond to
the overall position of the configuration in $AdS_5\times{\Bbb R}P^5$ 
along with the fermionic collective coordinates corresponding to the 
16 supersymmetric and superconformal fermion zero modes. The remaining 
non-abelian $\SU(k)$ partition function can then be explicitly
computed to give a factor proportional to $\sum_{d|k}d^{-2}$, a sum
over over the positive integer divisors $d$ of $k$. In addition to
this there is an overall numerical factor
$\sqrt N g^8 k^{-7/2}$ identical to the $\SU(N)$ case.

These relations imply that instanton contribution at charge
$2k$ and $k$, respectively, to various 
correlation function in the $\Sp(N)$ and $\SO(N)$ theories match
precisely the charge $k$ instanton contributions to the same 
correlators in the $\SU(N)$ theory.

\section{The ADHM Formalism for the Classical Groups}

The ADHM formalism \cite{ADHM} for constructing instanton solutions was 
adapted for dealing with arbitrary classical gauge groups
in the early instanton literature. 
The method adopted was to consider the construction for
one of the series of classical groups, e.g.~symplectic groups in
Ref.~\cite{Corrigan:1978ce} and 
orthogonal groups in Ref.~\cite{Christ:1978jy}, and then embed the other two series
in this series. Our approach will be no different, although we will
start from the unitary series. This has the advantage of avoiding the
language of quaterions but naturally our construction will be equivalent to those in
the previously mentioned references.

After briefly reviewing the ADHM formalism an $\N=4$ gauge theory with
gauge group $\SU(N)$, we
will then show how the orthogonal and symplectic cases may be embedded
therein. Full details of the $\SU(N)$ can be found in \cite{KMS,MO-III}.

The instanton solution at charge $k$, is
described by an $(N+2k)\times2k$ dimensional matrix $a$, and its
conjugate $\bar a$, with the form
\begin{equation}
a=\begin{pmatrix} w_\aD \\ a'_{\alpha\aD} \end{pmatrix}\ ,\qquad
\bar a=\begin{pmatrix} \bar w^\aD & \bar a^{\prime\aD\alpha} \end{pmatrix}\ .
\end{equation}
Here $w_\aD$ is a (spacetime) Weyl-spinor-valued $N\times k$ matrix and
$a'_{\alpha\aD}=a'_n\sigma^n_{\alpha\aD}$ where $a'_n$ is a (spacetime)
vector-valued $k\times k$ 
matrix. The conjugates are defined as
\begin{equation}
\bar w^\aD\equiv(w_\aD)^\dagger\ ,\qquad \bar
a^{\prime\aD\alpha}\equiv(a'_{\alpha\aD})^\dagger\ .
\label{hermcond}\end{equation}
The matrices $a'_n$ are
restricted to be hermitian: $(a'_n)^\dagger=a'_n$.
The remaining $4k(N+k)$ variables are still an over-parametrization of
the instanton moduli space which is obtained by a hyper-K\"ahler
quotient construction. One first imposes the ADHM constraints:
\begin{equation}
D^\aD_{\ \bD}\equiv\bar w^\aD w_\bD+\bar a^{\prime\aD\alpha}a'_{\alpha\bD}=\lambda
\delta^{\aD}_{\ \bD}\ ,
\label{badhm}\end{equation}
where $\lambda$ is an arbitrary constant matrix. The ADHM moduli space is then
identified with the space of $a$'s subject to \eqref{badhm} modulo the
action of an $\U(k)$ symmetry which acts on the instanton indices of
the variables as follows
\begin{equation}
w_\aD\rightarrow w_\aD U\ ,\qquad a'_{\alpha\aD}\rightarrow
U^\dagger a'_{\alpha\aD}U\ ,\quad U\in\U(k).
\end{equation}

The final piece of the story is the explicit construction of the
self-dual gauge field itself. To this end we define the matrix 
\begin{equation}
\Delta(x)=\begin{pmatrix}w_\aD \\ x_{\alpha\aD}1_{\sst[k]\times[k]}+
a'_{\alpha\aD}\end{pmatrix}\ ,
\end{equation}
where $x_{\alpha\aD}$, or equivalently $x_n$, related via
$x_{\alpha\aD}=x_n\sigma^n_{\alpha\aD}$, is a point in spacetime.
For generic $x$, the $(N+2k)\times N$ dimensional complex-valued matrix $U(x)$
is a basis for ${\rm ker}(\bar\Delta)$:
\begin{equation}
\bar\Delta U= 0 = \bar U\Delta\ ,
\label{uan}\end{equation}
where $U$ is orthonormalized according to
\begin{equation}
\bar U U = \ 1_{{\sst[N]}\times{\sst [N]}}\,.
\label{udef}\end{equation}
The self-dual gauge field is then simply
\begin{equation}v_n = 
\bar U \partial_{n}U\ .
\label{vdef}\end{equation}

The fermions in the background of the instanton lead to new Grassmann
collective coordinates. Associated to the fermions in the adjoint and
anti-symmetric tensor representations, $\lambda^A$, these collective
coordinates are described by the $(N+2k)\times k$ matrices $\M^A$,
and their conjugates\footnote{Here
$A=1,2,3,4$ is the spinor index of the $\SU(4)_R$ symmetry.}
\begin{equation}
\M^A=\begin{pmatrix} \mu^A \\ \M^{\prime A}_\alpha \end{pmatrix}\
,\qquad
\bar\M^A=\begin{pmatrix} \bar\mu^A & \bar\M^{\prime\alpha A} \end{pmatrix}\ ,
\end{equation}
where $\mu^A$ are $N\times k$ matrices and $\M^{\prime A}_\alpha$ are Weyl-spinor-valued
$k\times k$ matrices. The conjugates are defined by
\begin{equation}
\bar\mu^A\equiv(\mu^A)^\dagger\ ,\qquad
\bar\M^{\prime\alpha A}\equiv(\M^{\prime A}_\alpha)^\dagger\ ,
\end{equation}
and the constraint $\bar\M^{\prime A}_\alpha=\M^{\prime A}_\alpha$ is
imposed. 
These fermionic collective coordinates are subject to analogues of the
ADHM constraints \eqref{badhm}:
\begin{equation}
\lambda^A_\aD\equiv\bar w_\aD\mu^A+\bar\mu^Aw_\aD+[\M^{\prime\alpha
A},a'_{\alpha\aD}]=0\ .
\label{fadhm}\end{equation}

After this brief description of the $\SU(N)$ ADHM formalism we now turn
to the other classical groups $\Sp(N)$ and $\SO(N)$.\footnote{For the
orthogonal groups we restrict $N\geq4$.} In order to construct instanton
solutions for gauge theories with these groups we can use the embeddings
\begin{equation}
\Sp(N)\subset \SU(2N)\ ,\qquad \SO(N)\subset \SU(N)\ ,
\end{equation}
to extract the ADHM formalism for these groups in terms of the
$\SU(N)$ ADHM construction. The surprising feature of the resulting formalisms
is that the auxiliary group, $\U(k)$, in the $\SU(N)$
case and denoted generally as $\H(k)$, at instanton number $k$, in the general case
is {\it not\/} in the same classical series as the gauge group
$G$. Table 1 shows the
auxiliary groups and defines the quantities $N'$ and $k'$
which allow us to present a unified treatment of $\Sp(N)$ and $\SO(N)$.

\begin{center}
\begin{tabular}{cccc}
\hline
\Rowspace $G$ & $\SU(N)$ & $\Sp(N)$ & $\SO(N)$\\
\Rowspace $\H(k)$ & $\U(k)$ & $\O(k)$ & $\Sp(k)$\\ 
\Rowspace $ N'$ & $N$ & $2N$ & $N$\\ 
\Rowspace $ k'$ & $k$ & $k$ & $2k$\\ 
\hline 
\end{tabular}\\ 
\vspace{0.5cm}
Table 1: Gauge and associated auxiliary groups
\end{center} 

To describe the other classical groups we start with the theory with
gauge group $\SU( N')$ at instanton instanton number 
$k'$. Instanton solutions in the $\Sp(N)$ and $\SO(N)$ 
theories follow by simply imposing certain reality conditions on the
ADHM construction of the $\SU( N')$ theory which ensures that the
gauge field lies in the correct $sp(N)$ and $so(N)$
subalgebra of $su( N')$.
In order to deal with both the $\Sp(N)$ and $\SO(N)$ case at the same time
it is useful to define the notion of a generalized transpose operation
denoted $t$ which acts either on gauge or instanton indices.
Specifically, on $\Sp(n)$ group indices $t$ acts as 
as a symplectic transpose, i.e. on a column vector
$v$, $v^t=v^TJ^T$, where $J$ is $2n\times2n$ the symplectic matrix
\begin{equation}
J=\begin{pmatrix} 0 & 1 \\ -1 & 0\end{pmatrix}\ ;
\end{equation}
while on $\O(n)$ group indices $t$ is a conventional transpose
$t\equiv T$. The adjoint representations of both groups
are hermitian $t$-anti-symmetric matrices with dimensions $n(2n+1)$
and $n(n-1)/2$, respectively. Hermitian $t$-symmetric matrices
correspond to the {\it anti-symmetric\/} representation of $\Sp(n)$,
with dimension $n(2n-1)$ and the symmetric representation of $\SO(N)$,
with dimensions $n(n+1)/2$. For the symplectic groups $t$-(anti-)symmetric
matrices are conventionally called (anti-)self-dual \cite{Mehta}. 

The reality conditions on the bosonic collective coordinates
can then be written compactly as
\begin{equation}
\bar w^\aD=
\epsilon^{\aD\bD}(w_\bD)^t\ ,\qquad 
(a'_{\alpha\aD})^t=a'_{\alpha\aD}\ ,
\label{breality}\end{equation}
These reality conditions are only preserved by the subgroup
$\H(k)\subset\U(k')$ of the auxiliary symmetry group.
Given \eqref{hermcond} and \eqref{breality}, one can see that
the matrices $a'_n$ are hermitian and $t$-symmetric, i.e.~real 
symmetric in the case of auxiliary group $\O(k)$, and
symplectic anti-symmetric in the case of auxiliary group $\Sp(k)$. 
It is easy to verify that the
ADHM constraints \eqref{badhm} themselves are anti-hermitian
$t$-anti-symmetric, in other words $\H(k)$ adjoint-valued.
It is straightforward to show that these reality conditions are
precisely what is required to render the gauge field 
$t$-anti-symmetric, in other words to restrict it
to an $sp(N)$ and $so(N)$ subalgebra of $su(N')$.

The fermionic collective coordinates are subject to a similar set of reality
conditions: 
\begin{equation}
\bar\mu^A=(\mu^A)^t\ ,\qquad
\qquad(\M^{\prime A}_\alpha)^t=\M^{\prime A}_\alpha\ .
\label{freality}\end{equation}
So ${\cal M}^{\prime A}_\alpha$ is $t$-symmetric. The fermionic ADHM
constraints \eqref{fadhm} are, like their bosonic counterparts
$t$-anti-symmetric.\footnote{In proving this it 
is useful to notice that $(w_\aD^t)^t=-w_\aD$
and $((\mu^A)^t)^t=-\mu^A$.}

We can now count the number of real physical bosonic
and fermionic collective coordinates. For both $\Sp(N)$ and $\SO(N)$ at
instanton number $k$ there are 
$4kN$ independent $w$ variables, taking into account
the reality conditions. The number
of $a'_n$ variables is 
$4\times k(k+1)/2$ and $4\times k(2k-1)$, for $\Sp(N)$ and $\SO(N)$,
respectively. The physical moduli space is then the space of these
variables modulo the three $\H(k)$-valued 
ADHM constraints \eqref{badhm} and auxiliary $\H(k)$ symmetry.
Hence the dimension of the physical moduli space is
$4k(N+1)$ and $4k(N-2)$, for $\Sp(N)$ and $\SO(N)$, respectively. This
agrees with the counting via the index theorem.
The counting of the fermionic sector of the physical moduli space 
goes as follows. For each $A$, there are $2kN$ real
degrees-of-freedom in $\mu^A$ and $2\times k(k+1)/2$
and $2\times k(2k-1)$, in ${\cal
M}^{\prime A}_\alpha$, for $\Sp(N)$ and $\SO(N)$, respectively.
The ADHM constraints then impose $2\times k(k-1)/2$
and $2\times k(2k+1)$ conditions, for $\Sp(N)$ and $\SO(N)$,
respectively. Hence there are $2k(N+1)$ and $2k(N-2)$ real physical   
fermionic collective coordinates for $\Sp(N)$ and $\SO(N)$, respectively,
for each $A$. Again this agrees with the counting via the index theorem.

\section{The Multi-Instanton Collective Coordinate Measure}

In this section we write down the measure on the space of ADHM
collective coordinates and then show how, for $N$ large enough, both
the bosonic and fermionic ADHM constraints can be explicitly
resolved. This leads to a gauge invariant form for the measure which
is then amenable to a large-$N$ limit.

\subsection{The ADHM multi-instanton `flat' measure}

In order to calculate physical quantities we need to know how to
integrate on the space of ADHM variables. This is the measure induced from
the full functional integral of the field theory. Thankfully, it turns out the
measure is remarkably simple when written in terms of the 
ADHM variables: it is just the flat measure for all
the variables with the ADHM constraints imposed via explicit delta
functions \cite{meas1,meas2,MO-III}. In order to define the physical measure, we must divide by
the volume of the auxiliary group $\H(k)$:
\begin{equation}\begin{split}
\int d\mu_{\rm phys}^k=&
{a_{k}\left(C^{\prime\prime}_1\right)^{k} \over {\rm Vol}\ \H(k)}
\int da'\, dw\,\prod_{A=1,2,3,4}d{\cal M}^{\prime A}\, d\mu^A\,\prod_{B=2,3,4}
\ d\A^{1B}\\
&
\qquad\qquad\times
\prod_{c=1,2,3}\delta\big((\tau^c)^\bD_{\ \aD}
D^\aD_{\ \bD}\big)\prod_{A=1,2,3,4}\delta\big(\lambda^A_\aD\big)
\prod_{B=2,3,4}
\delta\big(\BL\cdot\A^{1B} -
\Lambda^{1B}\big)\ .
\label{measure}\end{split}\end{equation}
The form of the measure is an obvious generalization of the $\SU(N)$
measure \cite{MO-III,meas2} to the other groups. The following points are worthy of mention: 

(i) The integrals over the $t$-symmetric hermitian matrices $a'_n$ and
${\cal M}^{\prime A}_\alpha$ are defined as the integrals over the
components with respect to a basis $R^r$ of hermitian $t$-symmetric
matrices, normalized so that ${\rm tr}_{ k'}\,R^rR^s=\delta^{rs}$.

(ii) The delta functions for the ADHM constraints
and the integrals over the pseudo collective coordinates $\A^{AB}$ are defined
with respect to a basis of $t$-anti-symmetric
hermitian matrices, i.e.~the Lie algebra of $\H(k)$, normalized so that ${\rm
tr}_{ k'}\,T^rT^s=\delta^{rs}$.

(iii) The constant factor $a_{k}$ is fixed by clustering decomposition 
as discussed in the Appendix. Up to a factor of $c^k$, which can
absorbed into $(C_1^{\prime\prime})^k$, we prove
\begin{equation}
a_{k}=\begin{cases}2^{-k^2/4} & \Sp(N)\ , \\ 2^{-2k^2} & \SO(N)\ .
\end{cases} \label{cconst}
\end{equation}

(iv) The factor involving the $k^{\rm th}$ power of the constant
$C_1^{\prime\prime}$ is not fixed by clustering. It can be fixed by a
comparison with the explicit one instanton measure that can be deduced
in a similar way to that of the one instanton $\SU(N)$ measure in
\cite{Bernard:1979qt} to yield 
\begin{equation}
C_1^{\prime\prime}=(a_1)^{-1}\times\begin{cases}\big({g^2\over2\pi^3}\big)^{2(N+1)}
& \Sp(N)\\
\big({g^2\over2\pi^3}\big)^{2(N-2)} & \SO(N)\ .\end{cases}
\end{equation}

(v) The pseudo collective coordinates $\A^{AB}$ are $t$-anti-symmetric,
or $\H(k)$ Lie algebra-valued,
matrices associated to the scalar fields of the theory.
The operator $\BL$ is a linear operator on the Lie algebra of $\H(k)$ with a
positive spectrum, defined by
\begin{equation}
\BL\cdot\Omega={1\over2}\{\Omega,W^0\}+[a'_n,[a'_n,\Omega]]\ .
\label{E35}\end{equation}
The integrals over the $\A^{AB}$ can be done explicitly to yield
\begin{equation}
\int\prod_{B=2,3,4}d\A^{1B} \
\prod_{B=2,3,4}
\delta\big(\BL\cdot\A^{1B} - \Lambda^{1B}\big)
=\big({\det}\,\BL\big)^{-3}\ .
\label{trivint}\end{equation}
However, the un-integrated form \eqref{measure} makes the supersymmetry
more manifest \cite{meas2}.

\subsection{The multi-instanton action}

The action for the theory evaluated on the instanton solution at
leading order has the form
\begin{equation}
\Skinst\ =\ {8\pi^2k\over g^2}\ -ik\theta\ +\ S^k_{\rm quad}\ .
\label{Skinstdef}\end{equation}
Here $\Skquad$ is a particular fermion quadrilinear term, with one
fermion collective coordinate chosen from each of the four gaugino
sectors $A=1,2,3,4\,$ \cite{MO-III}:
 \begin{equation}\Skquad\ =\ 
{\pi^2\over
g^2}\,\epsilon_{ABCD}\,{\rm tr}_k\,\Lambda^{AB}
{\cal A}^{CD}\
 =\ {\pi^2\over
g^2}\,\epsilon_{ABCD}\,{\rm tr}_k\,\Lambda^{AB}
\BL^{-1}\Lambda^{CD}\  .  
\label{Skquadef}\end{equation}
Here 
\def\Lambdabar{\bar\Lambda}
\begin{equation}\Lambda^{AB}\ =\ {1\over2\sqrtwo}\,
\big(\,\Mbar^A\M^B -\Mbar^B\M^A \,\big)\ .
\label{newmatdef}\end{equation}
It is straightforward to show that 
$\Lambda_{AB}$ is $\H(k)$ adjoint-valued, i.e.~$t$-anti-symmetric.\footnote{In order to
show that $\Lambda_{AB}$ is $t$-anti-symmetric, it is important to
bear in mind that $t$ just acts on gauge and instanton indices. Hence
on a product of Grassmann quantities, say $A$ and $B$, there is an
extra minus sign under $t$-conjugation: $(AB)^t=-B^tA^t$.} 

As in \cite{MO-III}, the four-fermion interaction can be bi-linearized by introducing an
$\SO(6)$ $R$-symmetry vector
of hermitian $t$-anti-symmetric matrices $\chi_a$ (i.e.~$\H(k)$
adjoint-valued), $a=1,\ldots,6$. We can re-write $\chi_a$ as an
anti-symmetric tensor by introducing the Clebsch-Gordon coefficients
$\Sigma^a_{AB}$ (see \cite{MO-III} for definitions):
\begin{equation}  
\chi_{AB}={1\over\sqrt8}\Sigma^a_{AB}\chi_a\ .
\label{E54.1}\end{equation}
The identity we need is\footnote{In this and subsequent equations the
upper signs will correspond to gauge group $\Sp(N)$ and the lower signs
to gauge group $\SO(N)$.}
\begin{equation}
(\det\,\BL)^{-3}\exp\left(-S_{\rm quad}^k\right)
=\pi^{-3 k'( k'\mp1)}\int
d\chi\,\exp\big[-{\rm tr}_{ k'}\,\chi_a\BL
\chi_a+4\pi ig^{-1}{\rm
tr}_{ k'}\,\chi_{AB}\Lambda^{AB}\big]\ .
\label{E53}\end{equation}
Here, the integral is defined with respect to the $t$-anti-symmetric
basis: $\chi_a=\chi_a^rT^r$. Notice that this bi-linearization
absorbs the $(\det\,\BL)^{-3}$ factor from the $\A^{AB}$
integrals \eqref{trivint}.

\subsection{The gauge invariant measure}

It is convenient to change variables to a gauge invariant set. Since
we will be integrating gauge invariant quantities, the integral over
gauge transformation yields a volume factor. The gauge invariant
variables are encoded in the $2 k'\times2 k'$ matrix
$W$:
\begin{equation}
W_{\ \,\bD}^{\aD}=\bar w^\aD w_{\bD}\ .
\label{E25}\end{equation}
Associated to this are four $ k'\times k'$ hermitian matrices defined via
\begin{equation}
W^0={\rm tr}_2\,W,\quad W^c={\rm
tr}_2\,\tau^cW, \ \ c=1,2,3\ .
\label{Wdef}\end{equation}
Using the reality conditions \eqref{breality} one finds that in
addition to being hermitian
\begin{equation}
(W^0)^t=W^0\ ,\qquad (W^c)^t=-W^c\ .
\end{equation}
The bosonic ADHM \eqref{badhm} constraints are then linear in $W^c$:
\begin{equation}
0=W^c+i[a'_n,a'_m]{\rm tr}_2\,\tau^c\bar\sigma^{nm}=W^c-i[a'_n,a'_m]
\bar\eta^c_{nm}\ ,
\label{adhmw}\end{equation}
where $\bar\eta^c_{nm}$ is a 't~Hooft eta-symbol \cite{MO-III}.

When $ N'\geq2 k'$, we can change variables from the $w$'s to
the $W$'s yielding a Jacobian
as well as a numerical factor that reflects the volume of
the gauge group action:
\begin{equation} 
\int_{\rm gauge\atop coset}dw\,=\, c_{k,N}\int
\left({\rm
det}_{2 k'}W\right)^{ N'/2- {k'}^2/2\pm1/4} \,
dW^0\,\prod_{c=1,2,3}dW^c\ .
\label{E36}\end{equation} 
We will only need the numerical pre-factor in the large $N$
limit which can be evaluated by integrating both sides of \eqref{E36}
again a suitable exponential test function to give 
\begin{equation}
c_{k,N}\underset{N\rightarrow\infty}=2^{-2 {k'}^2\pm k'}( N'/\pi)^{-2kN+
k^{\prime2}\mp k'/2}e^{2kN}\ .
\end{equation}
This change of variables brings with it a significant advantage: we can
integrate out the $W^c$ variables using the delta functions
in \eqref{measure} since the latter are linear in the former.

On the fermionic side we need to integrate out the superpartners of the
gauge degrees-of-freedom. To isolate these variables we expand
\begin{equation}
\mu^A=w_{\aD}\zeta^{\aD A}+w_\aD\sigma^{\aD A}+\nu^A\ ,
\label{E44}\end{equation}
where the $\nu^A$ modes---the ones we need to integrate out---are 
in the subspace orthogonal to the
$w$'s, i.e.~$\bar w^\aD\nu^A=0$, and the variables $\zeta^{\aD A}$ and $\sigma^{\aD A}$ are
$t$-symmetric and $t$-anti-symmetric $ k'\times k'$ matrices, respectively. 
Notice that the $\nu^A$ modes do not appear in the fermionic
ADHM constraints \eqref{fadhm}. The change of variables from $\mu^A$ to
$\{\zeta^A,\sigma^A,\nu^A\}$ involves a Jacobian:
\begin{equation}
\int\, \prod_{A=1,2,3,4}d\mu^A=\int\left(\det_{2 k'}W\right)^{-2\hat
k}\,\prod_{A=1,2,3,4}
d\zeta^A\,d\sigma^A\,d\nu^A\ . 
\end{equation}
We can now integrate out the $\nu^A$ variables:
\begin{equation}
\int\, \prod_{A=1,2,3,4} d\nu^A\, 
\exp\big[\sqrt8\pi ig^{-1}
{\rm tr}_{ k'}\,\chi_{AB}(\nu^A)^t\nu^B\big]=2^{6kN
-6 {k'}^2}(\pi/g)^{4kN-4 {k'}^2}
\left({\rm det}_{4 k'}\chi\right)^{ N'/2- k'}\ .
\label{E55}\end{equation}
Furthermore, the fermionic ADHM constraints can then be used to integrate out the
$t$-anti-symmetric variables $\sigma^{\aD A}$, however the result is
rather cumbersome to write down and is in any case is not needed at
this stage. Fortunately in the large-$N$ limit to be described shortly
it simplifies considerably.

\section{The Measure in the Large-$N$ Limit}

The gauge invariant form of the measure is now
amenable to a large $N$
limit. As in \cite{MO-III}, we write terms which have the form of 
``something to the power $N$'' as 
$\exp(- N' S_{\rm eff}/2)$, for some ``effective large-$N$ action'' $S_{\rm eff}$,
and then perform a saddle-point approximation by minimizing the action
with respect to the variables. In order to achieve this,
we have to re-scale the $\chi$ variables:
\begin{equation} 
\chi_a\rightarrow\sqrt{ N'/2}\,\chi_a.
\label{E56}\end{equation}
The effective action has three contributions
\begin{equation}S_{\rm
eff}=-\log{\rm det}_{2 k'}\,W-\log{\rm det}_{4 k'}\,\chi
+{\rm tr}_{ k'}\,\chi_a\BL\chi_a\, ,
\label{E57}\end{equation}
up to constant which we keep track of separately.

The resulting saddle-point equations are identical to those in \cite{MO-III} for
the $\SU(N)$ case:
\begin{subequations}
\begin{align}
\epsilon^{ABCD}\left(\BL\chi_{AB}\right)
\chi_{CE}&=\tfrac12\delta_{E}^D1_{\sst [ k']\times[ k']}\ ,\label{E58}\\
\chi_a\chi_a&=\tfrac12(W^{-1})^0\ ,\label{E59}\\
[\chi_a,[\chi_a,a'_n]]&=i\bar\eta^c_{nm}[a'_m,(W^{-1})^c]\ .
\label{E60}
\end{align}
\end{subequations}

Using the $\SU(N)$ case \cite{MO-III} as a guide, we can write down
the solutions to these equations. 
As in the $\N=4$ case we look for a solution with $W^c=0$,
$c=1,2,3$, which means that the instantons are embedded in mutually commuting
$\SU(2)$ subgroups of the gauge group. In this case 
equations \eqref{E60} are equivalent to
\begin{equation}
[a'_n,a'_m]=[a'_n,\chi_a]=[\chi_a,\chi_b]=0\ ,\quad
W^0=\tfrac12(\chi_a\chi_a)^{-1}\ .\label{echo}
\end{equation}
The final equation can be viewed as giving the value of $W^0$, whose
eigenvalues are the
instanton sizes at the saddle-point. Clearly $\chi_a\chi_a$ and
$W^0$ must be non-degenerate.

For $\Sp(N)$ it is easy to see that a well-defined solution only exists for $k$
even. The point is that the set of commuting anti-symmetric matrices $\chi_a$
of odd dimension have a common null eigenvector and so $\chi_a\chi_a$
has no inverse contrary to hypothesis.\footnote{For $k$ odd one of the
eigenvalues of $W^0$ has to be infinite and although the action is
formally finite the $N$-independent
pre-factor in the measure is zero. We do investigate this rather
pathological solution in any more detail.}
{}From now on, for $\Sp(N)$ we shift $k\rightarrow2k$ and so $k'=k$
for both $\Sp(N)$ and $\SO(N)$ and we will replace $k'$ with $k$.
The general solution with instanton number $2k$, up to action of the
auxiliary $\O(2k)$ symmetry, is in block form
\begin{equation}\begin{split}
W^0&={\rm
diag}\big(2\rho_1,\ldots,2\rho_{k}\big)\otimes1_{\sst[2]\times[2]}\ ,\\
a'_n&={\rm
diag}\big(-X_n^1,\ldots,-X_n^{k}\big)\otimes1_{\sst[2]\times[2]}\ ,\\
\chi_a&={\rm diag}\big(\rho_1^{-1}\sfc_a^1,\ldots,\rho_{k}^{-1}
\sfc_a^{k}\big)\otimes\tau^2\ .
\end{split}\end{equation}
In the above the $\sfc_a^i$ are unit $SO(6)$ vectors.
The form of this solution is preserved by the following discrete
element of the auxiliary group:
\begin{equation}
1_{\sst[k]\times[k]}\otimes\tau^1\in\O(2k)\ ,
\end{equation}
which has the effect of changing the sign of $\sfc_a^i$. In other
words $\{\rho_i,X_n^i,\sfc_a^i\}$ parametrizes the positions of $k$
point-like objects, to be identified with the D-instantons of the dual
string theory, in $AdS_5\times S^5/{\Bbb Z}_2$ where the ${\Bbb Z}_2$ 
acts by inversion. This latter space is nothing
but the five-dimensional projective space ${\Bbb R}P^5$.

For $\SO(N)$ a solution exists for all instanton numbers and, up to the
auxiliary $\Sp(k)$ symmetry, is of the block form 
\begin{equation}\begin{split}
W^0&={\rm
diag}\big(2\rho_1,\ldots,2\rho_{k}\big)\otimes1_{\sst[2]\times[2]}\ ,\\
a'_n&={\rm
diag}\big(-X_n^1,\ldots,-X_n^{k}\big)\otimes1_{\sst[2]\times[2]}\ ,\\
\chi_a&={\rm diag}\big(\rho_1^{-1}\sfc_a^1,\ldots,\rho_{k}^{-1}
\sfc_a^{k}\big)\otimes\tau^3\ .
\end{split}\end{equation}
As in the $\Sp(N)$ case there is a discrete transformation 
\begin{equation}
1_{\sst[k]\times[k]}\otimes i\tau^2\in\Sp(k)\ ,
\end{equation}
that fixes the form of the solution but reverses the signs of the unit
six-vectors $\sfc^i_a$. So the solution describes the position of $k$
D-instantons in $AdS_5\times{\Bbb R}P^5$.

In principle, we have to expand the effective action around the general solutions
written down in the last section to sufficient order 
to ensure that the fluctuation integrals converge. In
general because the Gaussian form has zeroes whenever two D-instantons
coincide one has to go to quartic order in the fluctuations. 
Fortunately, as explained in \cite{MO-III}, 
we do not need to expand about the most general solution to the
saddle-point equations to quartic order since this is equivalent to expanding
to the same order around the most degenerate solution where all the 
D-instantons are at the same point in $AdS_5\times{\Bbb R}P^5$. 
The resulting quartic action has flat directions corresponding to the 
relative positions of the D-instantons. However, when the fermionic
integrals are taken into account the integrals over these relative
positions turn out to be convergent and hence these degrees-of-freedom
should be viewed as fluctuations around rather than facets of the maximally
degenerate solution. The variables left un-integrated, since they are
not convergent, can be viewed as centre-of-mass coordinates.

The maximally degenerate solution for both cases can be written as
\begin{equation}
W^0=2\rho^2\,1_{\sst[2k]\times[2k]}\ ,\qquad
\chi_a=\rho^{-1}\sfc_a\,S\ ,\qquad
a'_n=-\com_n\,1_{\sst[ 2k]\times[ 2k]}\ ,
\label{specsol}
\end{equation}
where
\begin{equation}
S=1_{\sst[k]\times[k]}\otimes\begin{cases}\tau^2 &
\Sp(N)\ , \\ \tau^3 & \SO(N)\ .\end{cases}
\end{equation}
The parameters $\{\rho,X_n,\sfc_a\}$ will be identified with centre-of-mass
coordinates. This solution is left invariant by a subgroup of $\U(k)\subset \H(k)$
consisting of transformations that fix $S$. For 
$\Sp(N)$ it is generated by the elements
\begin{equation}
{\rm Re}(a_{\sst[k]\times[k]})\otimes\tau^2\quad\text{and}\quad i\,{\rm Im
}(a_{\sst[k]\times[k]})\otimes1_{\sst[2]\times[2]} 
\label{ubssp}\end{equation}
and for $\SO(N)$ by elements
\begin{equation}
a_{\sst[k]\times[k]}\otimes\tau^3\ .
\label{ubso}\end{equation}
In both \eqref{ubssp} and \eqref{ubso} $a$ is a hermitian $k\times k$ matrix.

To construct the final form of the measure at large $N$, we have to expand the effective
action around the maximally degenerate solution to sufficient order to
ensure that the fluctuation integrals are convergent.
As in the $\SU(N)$ case \cite{MO-III}, some variables need only to
be expanded to Gaussian order whilst for the remainder, we must go to
quartic order. Although many details are identical to the $\SU(N)$
case in Ref.~\cite{MO-III} there are differences which we explain below:  

(i) As in \cite{MO-III}, we separate out the non-convergent integrals
over the centre-of-mass coordinates
$(\rho,X_n,\sfc_a)$ and expand the ADHM variables around the
maximally degenerate saddle-point solution as
\begin{subequations}
\begin{align}
a'_n&=-X_n1_{\sst [ 2k]\times[ 2k]}+\delta a'_n\ ,\\
W^0&=2\rho^21_{\sst [ 2k]\times[ 2k]}+\delta W^0\ ,\\
\chi_a&=\rho^{-1}\sfc_aS+\delta\chi_a\ ,
\end{align}\end{subequations}
It is useful to employ the further decomposition
\begin{equation}
\delta a'_n=\tilde a'_n+\hat  a'_n\ ,\qquad
\delta\chi_a=\tilde\chi_a+\hat\chi_a\ ,
\label{decom}\end{equation}
where $\tilde a'_n$ and $\tilde\chi_a$ are the fluctuations which do
not commute with $S$ (in fact they anti-commute)
and $\hat a'_n$ and $\hat\chi_a$ are the
remainder that do commute with $S$. 

Similarly, we separate out the integrals over the 16 supersymmetric
and super-conformal zero-modes $\xi^A_\alpha$ and $\bar\eta^{\aD A}$,
respectively, that are not lifted by the quadrilinear term
\eqref{Skquadef}, and expand the fermionic variables in a similar way
to the bosonic variables:
\begin{subequations}
\begin{align}
\M^{\prime A}_\alpha&=4\xi^A_\alpha1_{\sst [ 2k]\times[ 2k]}+\tilde\M^{\prime A}_\alpha
+\hat\M^{\prime A}_\alpha\ ,\\
\zeta^{\aD A}&=4\bar\eta^{\aD A}1_{\sst [ 2k]\times[ 2k]}+\tilde\zeta^{\aD
A}+\hat\zeta^{\aD A}\ ,
\end{align}
\end{subequations}
where we have introduced the same decomposition for the fermions as
for the bosons in Eq.~\eqref{decom}.

(ii) The expansion of the $\log\det_{8k}\chi$ term in the
saddle-point action can be read-off
from the 
$\SU(N)$ expression \cite{MO-III} with the following observation. In the present situation
$\chi$ is being expanded around $S$, rather than the identity in the
$\U(N)$ case, but since $S^2=1$ we can write
\begin{equation}
\chi_a=\rho^{-1}\sfc_aS+\delta\chi_a=S(\rho^{-1}\sfc_a1_{\sst[2
k]\times[2k]}+S\delta\chi_a)\ .
\end{equation}
Moreover $\det_{8k}\chi=\det_{8k}(S\chi)$ and so we can use
our previous $\SU(N)$ formulae for expanding $\det_{8k}\chi$
with the replacement $\delta\chi_a\rightarrow S\delta\chi_a$.

(iii) The expansion of the effective action to Gaussian order around
the saddle-point solution is 
\begin{equation}
S^{(2)}=-{1\over\rho^2}{\rm tr}_{2k}\,[S,\delta a'_n]^2-\rho^2
{\rm tr}_{2k}\,[S,\delta \chi_a^\perp]^2+
{\rm tr}_{2k}\,\big((2\rho^2)^{-1}\delta W^0+\rho\{S,\sfc_a\delta\chi_a\}\big)^2.
\end{equation}
In the above $\delta\chi_a^{\perp}$ is the component of $\delta\chi_a$
orthogonal to the unit vector $\sfc_a$. Writing the fluctuations in
terms of the decompositions \eqref{decom}, this becomes
\begin{equation}
S^{(2)}={4\over\rho^2}{\rm tr}_{2k}\,(\tilde a'_n)^2
+4\rho^2{\rm tr}_{2k}\,(\tilde \chi_a^\perp)^2+{\rm tr}_{2k}\,\varphi^2,
\label{gact}\end{equation}
where 
\begin{equation}
\varphi=2\{S,\sfc_a\delta\chi_a\}+{1\over2\rho^2}\delta W^0.
\end{equation}
Notice that the Gaussian terms lift all the fluctuations $\delta W^0$,
$\tilde a'_n$ and $\tilde\chi_a$ except the component of
$\tilde\chi_a$ along $\sfc_a$, which we denote
$\tilde\chi_a^\parallel$. In fact this variation corresponds precisely to the
action of infinitesimal $\H(k)$ transformations on the maximally
degenerate solution \eqref{specsol} although notice that as stated above the subgroup
$\U(k)$ leaves the solution invariant.
The observation is that any fluctuation $\tilde\chi_a^\parallel$ can be written as
$\sfc_a[S,\epsilon]$ for some $\epsilon\in \H(k)$ (the Lie algebra of $\H(k)$)
and conversely, infinitesimal $\H(k)$ transformations generate all such
$\tilde\chi_a^\parallel$. We can `gauge-fix' this symmetry by the
condition $\tilde\chi_a^\parallel=0$ which results in a Jacobian
factor that involves the volume of the coset
$\H(k)/\U(k)$.

Since ${\rm tr}_{2k}(\tilde x\hat y\hat z)=0$, for three
matrix quantities $x$, $y$ and $z$ with the decomposition \eqref{decom},
at leading order in $1/N$ the variables $\tilde a'_n$ and
$\tilde\chi_a$ are completely decoupled from the other fluctuations
and we can proceed to integrate them. Taking into account the
gauge-fixing Jacobian we have
\begin{equation}
\int d\tilde a'_n\,d\tilde\chi_a\,\exp\big(-2
N'{\rm tr}_{ 2k}\,[\rho^{-2}(\tilde
a'_n)^2+\rho^2(\tilde \chi_a^\perp)^2]\big)
=2^{9 k^2/2\mp k/2}(a'_{k})^{-1}{{\rm Vol}\,\H(k)\over{\rm Vol}\,\U( k)}
\Big({ N'\over\pi}\Big)^{-5 k^2}\ .
\end{equation}
Here the constant $a'_k$ is related to one defined in \eqref{cconst} with
$a'_{k}=a_{2k}$, for $\Sp(N)$, and $a'_k=a_k$, for $\SO(N)$.

(iv) The situations for the fermions is very similar. 
First of all we fulfill our
promise to deal with the fermionic ADHM constraints. To leading order
in $1/N$, these constraints read
\begin{equation}
2\rho^2\sigma_\aD^A=-\tfrac12[\delta W^0,\zeta_\aD^A]-[\M^{\prime\alpha
A},a'_{\alpha\aD}]\ .
\label{lnfadhm}\end{equation}
So the integrals over the $\sigma^{\aD A}$ variables soak up the
delta-functions imposing the fermionic ADHM constraints, as promised.
In \eqref{lnfadhm} $\delta W^0$ are the fluctuations in $W^0$ all of which are
lifted at Gaussian order. Due to a cross term we can effectively replace
$\delta W^0$ with $-4\rho^3\hat\Theta\cdot\chi$ at leading order (see
\cite{MO-III}). Collecting all the leading order terms, the fermion couplings
are
\begin{equation}\begin{split}
S_{\rm f}=i\Big({8\pi^2 N'\over g^2}\Big)^{1/2}
{\rm tr}_{2k}\Big[&
-2\rho^2(\hat\Theta\cdot\chi)\hat\Theta_{AB}
\zeta^{\aD A}\zeta_\aD^B+\rho^{-1}\hat\Theta_{AB}\big[a'_{\alpha\aD},
{\cal M}^{\prime\alpha A}\big]\zeta^{\aD B}\\
&\qquad\qquad\qquad+\chi_{AB}\big(\rho^2 
\zeta^{\aD A}\zeta_\aD^B+{\cal M}^{\prime
\alpha A}{\cal M}^{\prime B}_\alpha\big)\Big] \ .
\label{fermc}\end{split}\end{equation}
It is straightforward to see that the variables 
$\tilde\M^{\prime A}$ and $\tilde\zeta^A$ couple directly to the
saddle-point solution and can be integrated-out directly:
\begin{equation}\begin{split}
\int\,\prod_{A=1,2,3,4} d\tilde{\cal M}^{\prime A}\,d\tilde\zeta^A\,&\exp\big(
2 {N'}^{1/2}\pi g^{-1}\sfc_{AB}{\rm tr}_{2k}[\rho^{-1}S\tilde{\cal
M}^{\prime A}_\alpha\tilde{\cal M}^{\prime\alpha B}+\rho S\tilde\zeta^A_\aD
\tilde\zeta^{\aD B}]\big)\\
&\qquad\qquad\qquad=2^{ k^2\pm2 k}\Big({ N'\over g^2\pi^2}\Big)^{
k^2\pm2 k}\ .
\end{split}\end{equation}

(v) The remaining variables on the bosonic side are $\varphi$, $\hat
a'_n$ and $\hat\chi_a$. Fortunately the action for these variables,
to leading order is exactly as in the $\SU(N)$ case \cite{MO-III} with
the replacement, pointed out in (ii), $\hat\chi_a\rightarrow
S\hat\chi_a$. However, since all the hatted variables commute with $S$
in the final expressions this replacement is equivalent to
to leaving $\hat\chi_a$
unchanged and instead taking $\hat a'_n\rightarrow S\hat a'_n$,
$\hat{\cal M}^{\prime\alpha A}\rightarrow S\hat{\cal M}^{\prime\alpha
A}$ and $\hat\zeta^{\aD A}\rightarrow S\hat\zeta^{\aD A}$.
This is more convenient because then all the quantities
$\{S\hat a'_n,\hat\chi_a,S\hat{\cal M}^{\prime\alpha
A}, S\hat\zeta^{\aD A}\}$ are adjoint-valued in $\SU(
k)\subset \H(k)$. The variables $\varphi$ can be integrated out at
Gaussian order, as is evident from \eqref{gact}, however as in
\cite{MO-III} we have to take account of a coupling between $\varphi$
and a bi-linear in the remaining variables $\hat a'_n$ and
$\hat\chi_a$. Thus the $\varphi$ integral yields a quartic term in
these variables which along with the quartic terms in the expansions
of the two determinant factors in \eqref{E57} gives a remarkably simple
final result. As in \cite{MO-III} the remaining fluctuations can be 
assembled into a
ten-dimensional $\SU( k)$ gauge field with components
\begin{equation}
A_\mu=(N'/4)^{1/4}\big(\rho^{-1}S\hat a'_n,\rho\hat\chi_a\big),
\label{amudef}\end{equation}
and the quartic action for these variables is precisely
action of ten-dimensional $\SU(k)$ 
gauge theory dimensionally reduced to zero
dimensions:
\begin{equation}
NS_{\rm b}(A_\mu)\ =\ -{\rm tr}_{2k}\,\left[A_\mu,A_\nu\right]^2 \ .
\label{actymb}
\end{equation}

(vi) Not surprisingly the remaining fermionic variables
fermions can be assembled into a
Majorana-Weyl fermion of the dimensionally reduced ten-dimensional
theory with sixteen components
\begin{equation}
\Psi=\sqrt{\pi\over2g}( N'/4)^{1/8}\big(\rho^{-1/2}S\hat{\cal M}^{\prime A}_\alpha,
\rho^{1/2}S\hat\zeta^{\aD A}\big)\ .
\label{psidef}
\end{equation}
The coupling of these fermions to the gauge field $A_\mu$
completes the action \eqref{actymb} to that of a ten-dimensional
$\N=1$ supersymmetric $\SU(k)$ gauge theory dimensionally
reduced to zero dimensions:
\begin{equation}
S(A_\mu,\Psi)=-{\rm tr}_{2k}\,\left[A_\mu,A_\nu\right]^2 
+2{\rm
tr}_{2k}\,\bar\Psi\Gamma_\mu\left[A_\mu,\Psi\right]\, ,
\label{actym}
\end{equation}
where, in the way described in \cite{MO-III}, the ten-dimensional Gamma matrices
$\Gamma_\mu$ depend explicitly on the unit vector $\sfc_a$.

Our final result for
the measure in the large-$N$ limit is\footnote{In 
this result we have not kept track of factors of $2$ of the form
$2^{c_1+c_2k}$ ($c_1$ and $c_2$ being constants) since $2^{c_1}$
only affects the overall normalization of the answer and $2^{c_2k}$
can obviously be absorbed into a multiplicative redefinition of the modular
parameter $q=e^{2\pi i\tau}$; henceforth such a redefinition is understood.}
\begin{equation}
\int\dmuphys\,e^{-\Skinst}
\ \underset{N\rightarrow\infty}=\ 
{\sqrt{N}g^8e^{4\pi i k\tau}\over
 k^3\pi^{9 k^2/2+8}
{\rm Vol}\,\U( k)}\,
\int\,{d\rho\,d^4\com\over\rho^5}\, 
d^5\sfc\prod_{A=1,2,3,4}d^2\xi^A 
d^2\bar\eta^A\cdot\hat{\cal Z}_{k}\ .
\label{hello}\end{equation}
Here $\hat{\cal Z}_{ k}$ is the partition function of ten-dimensional
$\N=1$ supersymmetric $\SU( k)$ gauge theory dimensionally
reduced to zero dimensions:\footnote{Our normalization for the
partition is different from that in \cite{MO-III}.}
\begin{equation}
\hat{\cal Z}_k\ =\ \int_{\SU(k)}\, 
d^{10}A\, d^{16}\Psi\,e^{-S(A_\mu,\Psi)}\ .
\label{sukpart}
\end{equation}
To emphasize the above result pertains to
(i) $\Sp(N)$ in the sector with even instanton charge $2k$ and
(ii) $\SO(N)$ in all the charge sectors with charge $k$.
Notice that the form of the large-$N$ measure is identical
to that of of the charge $k$ sector of the $\SU(N)$ theory. 

At leading order in $1/N$ where any operator insertions take their
saddle-point values the $\SU( k)$ partition function  
$\hat{\cal Z}_{ k}$ is simply an overall constant
factor that has been evaluated in Ref.~\cite{KNS}:\footnote{We
have written the result in a way which allows an easy comparison with
\cite{KNS}. The factors of $\sqrt{2\pi}$ and $\sqrt2$ arise, respectively, from the
difference in the definition of the bosonic integrals and the normalization
of the generators: we have ${\rm tr}_k\,T^rT^s=\delta^{rs}$ rather
than $\tfrac12\delta^{rs}$. The
remaining factors are the result of \cite{KNS}.} 
\begin{equation}
\hat{\cal Z}_k=
\big(\sqrt{2\pi}\big)^{10(k^2-1)}\big(\sqrt2\big)^{(16-10)(k^2-1)}
\cdot
{2^{k(k+1)/2}\pi^{(k-1)/2}\over2\sqrt
k\prod_{i=1}^{k-1}i!}\cdot
\sum_{d\vert k}{1\over d^2}\ .
\label{parte}\end{equation}
Effectively for calculating correlation functions, we can take the
large-$N$ measure to be
\begin{equation}
\int\dmuphys\,e^{-\Skinst}
\ \underset{N\rightarrow\infty}{=}\ {\sqrt{ N}g^8\over
\pi^{27/2}}k^{-7/2}e^{4\pi i k
\tau}\sum_{d\vert k}{1\over d^2}\,
\int\,
{d^4\com\,d\rho\over\rho^5}\, d^5\sfc\prod_{A=1,2,3,4}d^2\xi^A 
d^2\bar\eta^A\ .
\label{endexp}
\end{equation}

To summarize, the large-$N$ instanton measures for the charge $2k$ sector of the
$\Sp(N)$ theory and the charge $k$ sector of the $\SO(N)$ theory, have
an identical form to that for the 
charge-$k$ sector in the $\SU(N)$ theory. So to leading order in
$1/N$, correlation functions will receive identical instanton
contributions from these sectors.

\acknowledgments

We acknowledge Matt Strassler for first pointing out to us the
generalization of the AdS/CFT correspondence to the other classical groups.
VVK and MPM acknowledge a NATO Collaborative Research Grant,
TJH and VVK acknowledge the TMR network grant FMRX-CT96-0012.

\startappendix

\rsen
\Appendix{Cluster Decomposition}

The clustering property of the $k$-instanton measure, namely the
degeneration of the measure into the $k-1$ and $1$ instanton measures
when one instanton is far separated from the others, is crucial for
determining the correct $k$ dependent normalization in front of the
measure (up to the constant $(C^{\prime\prime}_1)^k$). 
The clustering property has been extensively
discussed for the case of gauge groups $\Sp(1)\simeq \SU(2)$
\cite{meas1} and 
$\SU(N)$ in \cite{KMS}, so our discussion will be brief. 

We will consider the complete clustering limit which
describes a region of the moduli space where all the instantons are
well separated from each other. In this limit, we require
\begin{equation}
\int d\mu^k_{\rm phys}\ \longrightarrow\ 
\frac1{k!}\int d\mu^1_{\rm
phys}\times\cdots\times d\mu^1_{\rm
phys}\ .
\label{comclust}
\end{equation}
Intuitively, the complete clustering limit is the region of moduli
space where the matrices $[a'_n,a'_m]\approx0$ and where their
eigenvalues, which specify the positions of the individual instantons
are sufficiently part apart. We shall discover exactly what ``sufficiently
part apart'' actually means as we proceed.

It is useful to use the auxiliary $\H(k)$ symmetry to simultaneously
diagonalize the $t$-symmetric matrices 
$a'_n$ matrices in the clustering limit. This amounts
to ``gauge-fixing'' the coset 
$\H(k)/\H(1)^k$ of the auxiliary $\H(k)$
symmetry.\footnote{This gauge fixing is a valid procedure because the
measure is only used to integrate $\H(k)$ invariant functions.} The
remaining $\H(1)^{k}$ symmetry corresponds to the auxiliary symmetry groups
of the $k$ individual instantons.
The diagonal elements of $X^i_n=(a'_n)_{ii}$ then specify the positions
of the instantons (with $X^i=X^{i+k}$ for $\Sp(k)$). 
The gauge fixing involves a
Jacobian factor:
\begin{equation}
{1\over{\rm Vol}\,\H(k)}\int da'\ \longrightarrow\ 
\frac{c_k}{k!({\rm Vol}\,\H(1))^k}\int
\prod_{i=1}^kd^4X^i\,da^{\prime\perp}\,\prod_{1\leq i<j\leq k}
|X^i-X^j|^\beta\ .
\label{diagcov}
\end{equation}
where $\beta=1$, 2 and 4, for $\O(k)$, $\U(k)$ and $\Sp(k)$, respectively,
and the constant $c_k$ is 
\begin{equation}
c_k=\begin{cases}2^{k^2/4-3k/2} & \O(k)\ ,\\
2^{k(k+1)/2} & \U(k)\ ,\\
2^{2k^2-3k/2} & \Sp(k)\ ,\end{cases}
\end{equation}
and where \cite{Gilmore}
\begin{equation}
{\rm Vol}\,\H(k)=\prod_{j=1}^k\frac{2\pi^{j\beta/2}}{\Gamma(j\beta/2)}\ .
\end{equation}

In \eqref{diagcov} $a^{\prime\perp}_n$ are the off-diagonal components of 
$a'_n$ that are not generated by the adjoint action of $\H(k)$ on the
diagonal matrices ${\rm diag}(X^1_n,\ldots,X^k_n)$, for $\O(k)$ and
$\U(k)$, and ${\rm diag}(X^1_n,\ldots,X^k_n)\otimes 1_{\sst[2]\times[2]}$ for
$\Sp(k)$. Explicitly, these
are the projections
\begin{equation}
(a^{\prime\perp}_n)_{ij}=(a'_n)_{ij}-\frac{(X^i-X^j)_m(a'_m)_{ij}}{
|X^i-X^j|^2}(X^i-X^j)_n\ .
\end{equation}
The relation \eqref{diagcov} can easily be derived
from the well-known Jacobian that arises from changing variables for
an integral over the
elements of a (i) real symmetric (ii) Hermitian or a (iii) Hermitian self-dual
matrix $A$, for
$\O(k)$, $\U(k)$ and $\Sp(k)$, respectively (normalized with respect
to the basis $R^r$), to its eigenvalues
$a^i$ (see for example \cite{Mehta}):
\begin{equation}
\int dA=\frac{c_k}{k!}\frac{{\rm Vol}\,\H(k)}{({\rm Vol}\,\H(1))^k}
\int\prod_{i=1}^k da^i\,\prod_{1\leq i<j\leq k}|a^i-a^j|^\beta\
.
\label{meht}\end{equation}

The clustering limit is defined as the limit where
the off-diagonal bosonic ADHM constraints can be approximated by 
\begin{equation}
(X^i-X^j)^{\aD\alpha}(a^{\prime\perp}_{\alpha\bD})_{ij}+(\bar
w^\aD w_\bD)_{ij}=\lambda_{ij}\delta^\aD_{\ \bD}\ ,
\label{boscl}
\end{equation}
for arbitrary $\lambda_{ij}$,
i.e.~to order ${\cal O}(|X^i-X^j|^{-2})$ they are approximately linear in the elements
$(a^{\prime\perp}_n)_{ij}$. In this limit, we can then easily
integrate out these elements. Similarly, in the clustering limit, the
off-diagonal fermionic ADHM constraints are approximately linear in
the elements $(\M'_\alpha)_{ij}$; to order ${\cal O}(|X^i-X^j|^{-1})$
\begin{equation}
(X^i-X^j)_{\alpha\aD}(\M^{\prime\alpha})_{ij}
+(\bar\mu w_\aD)_{ij}+(\bar w_\aD\mu)_{ij}=0
\ .
\label{fermcl}
\end{equation}
In the clustering limit, we can then easily
integrate out the off-diagonal elements $(\M'_\alpha)_{ij}$.

After integrating-out the off-diagonal components of the ADHM
constraints using \eqref{boscl} and \eqref{fermcl}, the $k$
independent diagonal components
of the constraints are the ADHM constraints for the $k$ individual
instantons. Taking careful account of numerical factors one finds that
the measure in \eqref{measure} must include the factor $a_k$, defined
in \eqref{cconst} to cancel the factor of $2^{k^2}$ in
\eqref{diagcov}. Notice that the factor of $2^k$ can be absorbed into
the constant $(C^{\prime\prime}_1)^k$.

\end{document}